\documentclass[preprint,pdf]{iucr}              

\def\href#1{\relax}\let\foo\caption
\ifPDF
\RequirePackage{hyperref}
\PassOptionsToPackage{pdftex,bookmarksopen,bookmarksnumbered}{hyperref}
\fi
\let\caption\foo

     \paperprodcode{a000000}              
     \paperref{xx9999}                    
     \papertype{FA}                       

     \journalcode{S}              

\usepackage{amsmath} 
\usepackage{amsfonts}
\usepackage{algorithm}
\usepackage{algpseudocode}
\usepackage{setspace} 

\usepackage{amssymb}
\usepackage{graphicx} 
\usepackage{fixltx2e}

\usepackage[caption=false]{subfig}

\newcommand{\pic}[3]
{
\begin{figure}
\caption{#3}
\includegraphics[scale=#1]{#2}
\end{figure}
}

\newcommand{\argmin}[1]{\underset{#1}{\operatorname{argmin}} \; } 
\newcommand{\argmax}[1]{\underset{#1}{\operatorname{argmax}} \; } 

\renewcommand{\vec}{\boldsymbol} 
\newcommand{\tv}[1]{\operatorname{TV}\left( #1 \right)}
\newcommand{\prox}[2]{\operatorname{prox}_{#1}{\left(#2\right)}}

\newcommand{\sign}[1]{\operatorname{sign}\left( #1 \right)}

\renewcommand{\div}{\operatorname{div}}
\newcommand{\id}{\operatorname{Id}}
\newcommand{\norm}[1]{\left\| #1 \right\|}

\newcommand{\pmat}[1]{\begin{pmatrix} #1 \end{pmatrix}}
\newcommand{\E}[1]{\cdot 10^{#1}} 
\begin{document}                  



\title{Ring artifacts correction in compressed sensing tomographic reconstruction}
\shorttitle{Rings artifacts correction in compressed sensing tomographic reconstruction}


\cauthor[a,b]{Pierre}{Paleo}{pierre.paleo@esrf.fr}{}
\cauthor[a]{Alessandro}{Mirone}{mirone@esrf.fr}{}

\aff[a]{ESRF, 71 avenue des Martyrs, 38000 Grenoble \country{France}}
\aff[b]{Universit\'{e}  de Grenoble, Gipsa-Lab, 11 Rue des Math\'{e}matiques, 38400 Saint-Martin-d'H\`{e}res, \country{France}}


\shortauthor{Paleo and Mirone}







\maketitle                        


\begin{abstract}
We present a novel approach to handle ring artifacts correction in compressed sensing tomographic reconstruction. The correction is part of the reconstruction process, which differs from classical sinogram pre-processing and image post-processing techniques. 
The principle of compressed sensing tomographic reconstruction is presented. Then, we show that the ring artifacts correction can be integrated in the reconstruction problem formalism. We provide numerical results for both simulated and real data. 
This technique is included in the PyHST2 code which is used at the European Synchrotron Radiation Facility for tomographic reconstruction.

\end{abstract}

\section{Introduction}
\subsection{Rings artifacts in tomographic reconstruction}
During a tomographic acquisition process, some flaws in the experimental setup can lead to unwanted artifacts.
One instance is the presence of defective pixels in the detector, which appear as lines in the sinogram when the defects are independent of the projection angle. 
These spurious lines give raise to rings artifacts in the reconstructed object. 
Even after preprocessing steps like flat-field correction and median filtering, these artifacts can remain and are detrimental to the reconstruction quality.
Therefore, multiple techniques have been developed to tackle this problem.
\subsection{Related work}
Various techniques have been proposed in the literature to reduce or suppress the rings artifacts. As reported in \cite{ringscomparison}, these techniques can be classified into two groups : sinogram preprocessing and reconstructed images post-processing.
The preprocessing methods aims at detecting and correcting the spurious lines in the sinogram before applying the reconstruction process, thus, rings do not form if the method succeeds. 
On the other hand, post-processing techniques work directly on the reconstructed image, trying to extract the concentric circles and filter them.
These methods often perform a transformation into polar coordinates to transform the concentric circles into straight lines \cite{prell09}. 

A comprehensive comparison of ring artifact removal methods can be found in \cite{ringscomparison}. 
Although these methods certainly provide satisfactory results in their  limited framework, 
the authors report that no existing method is really suitable for correcting different types of rings, since they always introduce other distortions. 
A recent work \cite{miqueles} reports a compressed sensing approach for rings artifacts reduction using a total variation denoising of the sinogram before calling the reconstruction routine.
It is a generalization of Titarenko's Algorithm \cite{titarenkoReg} which consists in a regularization of the sinogram. This can also be classified in the sinogram pre-processing techniques.

\section{Preamble}
\textbf{\\} 
In this section, we introduce the principle and the formalism of compressed sensing tomographic reconstruction.
This formalism is extended in section \ref{ringscorr} for ring artifacts correction.

\subsection{Compressed sensing tomographic reconstruction}
Computed tomography aims at reconstructing an image $\vec{x}$ from a set of projections $\vec{y} = P(\vec{x})$. Here $\vec{y}$ denotes the acquired sinogram, $\vec{x}$ is the slice to be reconstructed and $P$ is the projection operator (while its adjoint $P^* = P^T$ is the back-projection operator).
The classical filtered backprojection algorithm 
enables to reconstruct the image, but the number of projections should be of the same order of the number of rows in the image to have an acceptable reconstruction according to the Shannon-Nyquist sampling theorem. 
This is often impracticable, and the subsampling leads to artifacts in the reconstructed image.

Compressed sensing techniques exploit a-priori knowledge on the image, like its sparsity, in order to bypass this limitation. 
Instead of computing a closed form solution like in the filtered backprojection technique, tomographic reconstruction by compressed sensing amounts to an optimization problem
\begin{equation}\label{generalpb}
\vec{\hat{x}} = \argmin{\vec{x}} f(\vec{y},\vec{x}) + g(\vec{x})
\end{equation}
where $f(\vec{y},\vec{x})$ is a fidelity term of $\vec{x}$ with respect to the acquired data $\vec{y}$ (henceforth $f(\vec{y},\vec{x})$ is denoted $f(\vec{x})$), and $g(\vec{x})$ contains a-priori knowledge on the image.
In general, the regularization term $g(\vec{x})$ makes the problem non-smooth, which precludes from using usual Gradient-like algorithms ${\vec{x}_{n+1} = \vec{x}_n - \gamma_n \nabla f (\vec{x}_n)}$.

Advances in convex analysis provide adapted methods, based on proximal splitting methods  \cite{proxsplitting}, which are a generalization of projected gradient. 
One instance is the Iterative Shrinkage-Thresholding Scheme (ISTA). For a functional split into a smooth term $f$ and a non-smooth term $g$, the first order condition at an optimum $\hat{x}$ reads
\begin{align*}
0 &\in \nabla f (\hat{x}) + \partial g (\hat{x}) \\
0 &\in \nabla f (\hat{x}) - \hat{x} + \hat{x} + \partial g (\hat{x}) \\
\left( \id + \partial g \right) (\hat{x}) &\in \left( \id - \nabla f \right) (\hat{x}) \\
\hat{x} &= \left(\id + \partial g\right)^{-1} \left(\id - \nabla f \right) (\hat{x})
\end{align*}
which suggests the use of a fixed point iterative scheme. The operator $\left(\id + \partial g\right)^{-1}$ is called \textit{proximal operator} :
\begin{equation}
\prox{\lambda g}{\vec{\hat{x}}} = \left( \id + \lambda \partial g \right)^{-1} (\hat{\vec{x}}) = \argmin{\vec{x}} \left\{ \dfrac{1}{2\lambda}\norm{\vec{x}-\vec{\hat{x}}}_2^2 + g(\vec{x}) \right\}
\end{equation}
so that one step of ISTA 
reads
\[
x_{k+1} = \prox{g/L}{x_k - \dfrac{1}{L} \nabla f(x_k)}
\]
where $L$ is the Lipschitz constant of the gradient $\nabla f$ of the fidelity term.
Usually, the data fidelity term is a L2 norm, so the calculation of the proximity operator of $f$ is straightforward. On the other hand, the proximity operator $\operatorname{prox}_{\lambda g}$ of the regularization term does not always have a closed form expression.

\subsection{Total variation regularization}\label{tvreg}
Depending on the regularization term used in \eqref{generalpb}, the reconstruction can yield very different results. If the regularization term is null, the problems amounts to a least-squares reconstruction. This approach tends to blur the edges in the reconstructed slice. 
A commonly used regularization term for image denoising is the Total Variation which was introduced in \cite{ROF} as the Rudin-Osher-Fatemi model. 
This prior has the property to preserve the image edges, which is essential especially in tomographic reconstruction where the object in the sample should be distinguishable. 
In a discrete framework, the Total Variation is the L1 norm of the gradient :
\begin{equation}
\tv{\vec{x}} = \norm{\nabla \vec{x}}_1 = \sum_i \sqrt{(\nabla_1 \vec{x})(i)^2 + (\nabla_2 \vec{x})(i)^2}
\end{equation}
Given an observed sinogram $\vec{y}$, the total variation tomography reconstruction problem aims at finding the regularized image $\vec{x}$ satisfying
\begin{equation}\label{tvtomo}
\argmin{\vec{x}} \left\{ \dfrac{1}{2} \norm{\vec{y} - P \vec{x}}^2  + \beta \tv{\vec{x}} \right\}
\end{equation}
where $\beta$ weights the regularization with respect to the data fidelity term, and $P$ is the projection matrix. 
If $P$ is equal to the identity matrix, the problem \eqref{tvtomo} amounts to the so-called total variation denoising problem :
\begin{equation}\label{tvdenoise}
\argmin{\vec{x}} \left\{ \dfrac{1}{2} \norm{\vec{y} - \vec{x}}_2^2  + \beta \tv{\vec{x}} \right\} = \prox{\beta\operatorname{TV}}{\vec{y}}
\end{equation}
In \cite{michel11}, a dual approach enables to compute the proximity operator of the total variation :
\begin{equation}
\left\{
\begin{array}{l}
\vec{\hat{z}} = \argmax{\norm{\vec{z}}_\infty \leq 1} \left\{ - \norm{\lambda\div \vec{z} + \vec{y}}_2^2 \right\} \\
\vec{\hat{x}} = \prox{\lambda \operatorname{TV}}{\vec{y}} = \vec{y} + \lambda \div \vec{\hat{z}}
\end{array}
\right.
\end{equation}
Knowing the proximal map of the Total Variation regularization term, the denoising problem can be solved by an iterative shrinkage-thresholding scheme. 
In PyHST2, we use an accelerated version known as the Nesterov algorithm \cite{nesterov} or FISTA \cite{beckteboulleIEEE}.

However, tomographic reconstruction is not a simple denoising problem, since a projection matrix $P$ appears in the problem \eqref{tvtomo}.
Constructing a smooth dual of \eqref{tvtomo} would entail to invert $P^T P$ which is an ill-posed problem. The total variation tomographic reconstruction is tackled with two nested FISTA loops, each iteration being the solution of a denoising subproblem \cite{beckteboulleIEEE}
\begin{equation}
\operatorname{denoise}\left( \vec{x} - \dfrac{1}{L} P^T \left( P \vec{x} - \vec{y}\right) \right)
\end{equation}
Here $L$ is the Lipschitz constant of $\nabla f (\vec{x}) = P^T (P \vec{x} - \vec{y})$ which is calculated with the power method.

\subsection{Dictionary learning}
Total Variation regularization performs well for piecewise-constant images since edges and uniform regions are reinforced. However, for non piecewise-constant images, the \textit{cartoon} effect might be prejudicial for the reconstruction quality. Thus, another regularization technique has to be considered for such images.

Most natural images have an intrinsic sparsity which can be recovered by an adapted transform 
or by building the best sparsifying basis. The latter technique is called dictionary learning. 
Given a set of $N$ acquired signals $\vec{y}_p$, a number of $N_c$ basis vectors (or atoms) $\vec{\varphi}_k$ are built. Each signal $\vec{y}_p$ is expressed as a linear combination $(w_{p,1}, \ldots, w_{p,N_c})$ of the $N_c$ atoms. Dictionary learning is a joint optimization of the atoms $D$ and the coefficients $\vec{w}_p$ under the constraint of sparsity :
\begin{equation}
\begin{aligned}
\argmin{D, W} \sum_p \norm{\vec{y}_p - D \vec{w}_p}^2 \\
\text{s.t.} \; \forall \, p \, , \; \norm{\vec{w}_p}_0 \leq S
\end{aligned}
\end{equation}
Here $\norm{\cdot}_0$ denotes the zero norm counting the number of non-zero component of a vector.
The dictionary $D$ is learned with K-SVD \cite{ksvd}.

In image processing problems like tomographic reconstruction, the image $\vec{x}$ is divided into $N$ square patches of size $m \times m$ pixels.
Every patch area of index $(p)$ is expressed as a linear combination of the atoms $\vec{\varphi}_k$ :
\begin{equation}
\vec{\operatorname{patch}}(p) = \sum_k w_{k,p} \vec{\varphi}_k
\end{equation}
That is, for pixel $\vec{i}$ of the image $\vec{x}$ :
\begin{equation}
\vec{x_i} = \sum_k w_{k,p_i} \varphi_k (\vec{i} - \vec{r}_{p_{\vec{i}}})
\end{equation}
where $p_{\vec{i}}$ denotes the patch containing the pixel $\vec{i}$, and $r_{p_{\vec{i}}}$ is the center of this patch so that $\vec{i}-r_{\vec{p_i}}$ belongs to the patch support.

To overcome discontinuities in patches borders, the PyHST2 \cite{pyhst2} approach allows the patches to overlap.
Following formalism used in PyHST2, $\vec{1_p(i)}$ denotes the indicator function of patch $(p)$ while $\vec{1_p^c (i)}$ denotes the indicator function of the patch center\footnote{each patch center is a subset of the patch such that the patch centers are not overlapping}. Every image pixel is then a linear combination of possibly overlapping atoms :
\begin{equation}
\vec{x_i} = \sum_p \vec{1_p^c(i)} \sum_k w_{k,p} \varphi_k (\vec{i} - \vec{r}_{p})
\end{equation}
The tomographic reconstruction problem is
\begin{equation}\label{pbdl1}
\argmin{W} \norm{\vec{y} - P \vec{x}(W)}^2 + \beta_{\text{DL}} \norm{\vec{w}}_1 + \rho \cdot h(W)
\end{equation}
where $W = \left( w_{k,p} \right)_{0 \leq k < N_c}^{0 \leq p < N}$ is the matrix containing the patches coefficients, and
\begin{equation}
h(W) = \sum_{p, \vec{i}} \vec{1_p (i)} \left( \vec{x_i} - \sum_k w_{k,p} \varphi_k (\vec{i}-\vec{r_p}) \right)^2
\end{equation}
is a term promoting patches overlapping. Again, the optimization problem \eqref{pbdl1} is solved with the FISTA algorithm.

\section{Rings correction in compressed sensing reconstruction}\label{ringscorr}

We now present how rings correction can be handled directly in the reconstruction process by integrating additional variables in the functional to minimize. This approach is independent of the regularization used and can therefore be applied in various frameworks like total variation and dictionary learning.
Sinogram pre-processing techniques modify the acquired data to filter the unwanted lines. This filtering often introduces new artifacts.
On the other hand, image correction techniques can also add new artifacts when circular features are detected as artifacts ; and the forward and backward Cartesian-polar coordinate transforms lead to a loss of precision even with a bilinear interpolation.
When the rings correction is done in the reconstruction process, the data is not modified ; it is an initial guess of the solution for the iterative method.
The solution minimizes the euclidean distance from the reference slice -- the backprojected sinogram -- under regularization constraints. The rings correction is included in both the constraints and the functional to minimize.

In our approach, the rings correction consists in splitting the sinogram into two components : the ``genuine" sinogram and the spurious straight lines giving rings after back-projection. This splitting is done in the reconstruction process, so the two components are updated after each iteration. In the end, only the valid sinogram component is kept while the rings variables are discarded. We give two examples of frameworks using this approach : total variation regularization and dictionary learning reconstruction.

\subsection{Rings correction in total variation framework}\label{ringstv}

Rings correction can be handled by additional variables $\vec{r}$ in the fidelity term $f(\vec{y},\vec{x})$.
These artifacts come from a spurious line along projection angles in the sinogram. 
Thus, rings variables are stacked in a vector and added to the sinogram for each projection. 
The fidelity term for one projection reads
\begin{equation}\label{fidelitytv}
f(\vec{x},\vec{r}) = \dfrac{1}{2} \norm{\vec{y} - (P \vec{x} + \vec{r})}_2^2
\end{equation}
here $P\vec{x}$ and $\vec{r}$ do not have the same dimensionality ; $P\vec{x}+\vec{r}$ means that a vector of rings variables is added to each line of the sinogram as illustrated on Figure \ref{fig:ringsvariables}.
The functional $F(\vec{x},\vec{r})$ is
\begin{equation}\label{ringstv}
F(\vec{x},\vec{r}) = f(\vec{x},\vec{r}) + \beta \tv{\vec{x}} + \beta_r \norm{\vec{r}}_1
\end{equation}
$\beta$ being a parameter weighting the relative importance of spatial regularization, and $\beta_r$ being a penalization parameter for the rings.

While sinogram preprocessing techniques filter the lines parallel to the projection angle, this approach forces the sinogram to be decomposed as a sinogram $P \vec{x}$ and rings variables $\vec{r}$. The sparsity constraint $\beta_r \norm{\vec{r}}_1$ forces the rings variables to have only a few not null components, since the L1 norm is a good approximation of the sparsity-inducing L0 norm.

We emphasize the fact that the sinogram decomposition into a genuine sinogram $P\vec{x}$ and spurious rings $\vec{r}$ is not a pre-processing technique ; the rings removal is intrinsically part of the reconstruction process. At each iteration, the image $\vec{x}$ and the rings variables $\vec{r}$ are adapted to minimize the energy $F(\vec{x},\vec{r}$).

\pic{0.4}{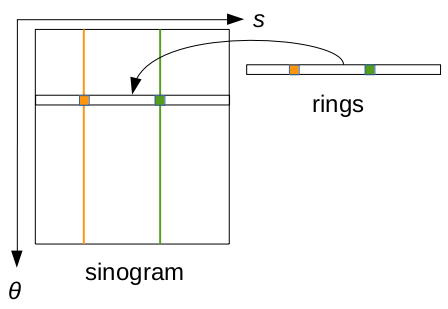}
{
\label{fig:ringsvariables}
Principle of the rings separation. $\theta$ is the projection angle and $s$ is the detector bin index.
The vertical orange and green lines represent spurious lines giving raise to ring artifacts.
The decomposition $P \vec{x} + \vec{r}$ forces the ring values to be captured in the vector $\vec{r}$ (independent of the projection angle).
In the end, only the part without the rings $\vec{r}$ is returned.
}


The minimum of $F(\vec{x},\vec{r})$ is found with the total variation regularization solver presented in \ref{tvreg}.
This iterative algorithm has a step size $\gamma = 1/L$ where $L$ is the Lipschitz constant of the gradient $\nabla f$.
This constant is an upper bound of the largest eigenvalue of the Hessian $\nabla^2 f$.
Since the gradient is now taken with respect to both image and rings variables, the Hessian is
\begin{equation}
\nabla_{\vec{x},\vec{r}}^2 f = 
\begin{bmatrix}
\nabla_{\vec{x}}^2 f & \nabla_{\vec{r},\vec{x}}^2 f \\
\nabla_{\vec{x},\vec{r}}^2 f & \nabla_{\vec{r}}^2 f
\end{bmatrix}
= \begin{bmatrix}
P^T P & P^T \\
P & \mathrm{Id}
\end{bmatrix}
\end{equation}
its largest eigenvalue is calculated with the power method.

Once the Lipschitz constant $L$ is obtained, one iteration of the FISTA denoising algorithm needs to compute
\begin{equation}\label{prox1}
\begin{split}
\vec{v}_{k+1} &= \prox{g/L}{\vec{v}_{k-1} - \dfrac{1}{L} \nabla f (\vec{v}_{k-1})} \\
 &= \argmin{\vec{z}} \left\{\dfrac{1}{2} \norm{\vec{z} - \left( \vec{v}_{k-1} - \dfrac{1}{L} \nabla f (\vec{v}_{k-1})\right)}_2^2 \right. \\
 &\qquad\qquad\qquad + \left.  \dfrac{1}{L} \beta \tv{\vec{z_x}} + \dfrac{1}{L} \beta_r \norm{\vec{z_r}}_1 \right\}
\end{split}
\end{equation}
Here $\vec{v}$ denotes the augmented vector $\vec{v} = \begin{pmatrix}\vec{x} \\ \vec{r}\end{pmatrix}$ containing both image and rings variables, $\vec{z_x}$ (resp. $\vec{z_r}$) is the part of vector $\vec{z}$ containing image (resp. rings) variables. 
Since the squared L2 norm is separable, the proximal operator \eqref{prox1} can be written
\begin{equation}\label{twoprox}
\begin{aligned}
\vec{v}_{k+1} &= \argmin{\vec{z}}  \Bigg\{ \dfrac{1}{2} \norm{\vec{x} - \left( \vec{x}_{k-1} - \dfrac{1}{L} \nabla f (\vec{x}_{k-1})\right)}_2^2 \\
 &\quad + \dfrac{1}{L} \beta \tv{\vec{z_x}} + \dfrac{1}{2} \norm{\vec{r} - \left( \vec{r}_{k-1} - \dfrac{1}{L} \nabla f (\vec{r}_{k-1})\right)}_2^2 \\
 &\quad + \dfrac{1}{L} \beta_r \norm{\vec{z_r}}_1 \Bigg\} \\
& = \prox{\beta\operatorname{TV}/L}{\vec{x}_{k-1}  - \dfrac{1}{L} \nabla_{\vec{x}} f (\vec{x}_{k-1})} \\ 
&\quad \oplus \prox{\beta_r\norm{\cdot}_1/L}{\vec{r}_{k-1} - \dfrac{1}{L} \nabla_{\vec{r}} f (\vec{r}_{k-1})} 
\end{aligned}
\end{equation}
In short, the proximal operator is separable with respect to the image and rings variables. This is convenient because rings and image variables can be updated by solving two separate subproblems in a FISTA iteration.

The first term in \eqref{twoprox} is the denoising problem \eqref{tvdenoise}. The second term is the proximity operator of the L1 norm, which has a closed-form expression called \textit{soft thresholding operator} :
\begin{equation}\label{proxl1}
\prox{\lambda \norm{\cdot}_1}{\vec{u}} = \max \left( |\vec{u}| - \lambda \, , \, 0 \right) \cdot \sign{\vec{u}} \qquad \text{element wise}
\end{equation}

\subsection{Rings correction in Dictionary Learning framework}\label{dltomo}
Similarly to total variation (\ref{ringstv}), the proximal operator is separable with respect to patch variables $W$ and rings variables $\vec{r}$. The problem \eqref{pbdl1} becomes
\begin{equation}
\argmin{W} \norm{\vec{y} - (P \vec{x}(W) +\vec{r})}_2^2 + \beta_{\text{DL}} \norm{\vec{w}}_1 + \beta_r \norm{\vec{r}}_1 + \rho \cdot h(W)
\end{equation}
On the other hand, in this case, the non-smooth term is only the L1 norm which has a closed-form proximal operator \eqref{proxl1}. Thus, the denoising problem is straightforward, and only one FISTA loop is required :
\begin{equation}
\begin{aligned}
\vec{x}_k &= \operatorname{soft}\_\operatorname{threshold}_{\beta_{\text{DL}}/L} \left( \tilde{\vec{x}}_k - \dfrac{1}{L} \nabla_W f(\tilde{\vec{x}}_k) \right) \\
\vec{r}_k &= \operatorname{soft}\_\operatorname{threshold}_{\beta_r/L} \left( \tilde{\vec{r}}_k - \dfrac{1}{L} \nabla_{\vec{r}} f(\tilde{\vec{r}}_k) \right) \\
t_{k+1} &= \dfrac{1+\sqrt{1+4t_k^2}}{2} \\
\tilde{\vec{x}}_{k+1} &= \vec{x}_k + \left( \dfrac{t_k -1}{t_{k+1}}\right) (\vec{x}_k - \vec{x}_{k-1}) \\
\tilde{\vec{r}}_{k+1} &= \vec{r}_k + \left( \dfrac{t_k -1}{t_{k+1}}\right) (\vec{r}_k - \vec{r}_{k-1})
\end{aligned}
\end{equation}

\subsection{Preconditioning the optimization problem}\label{precond}
In both cases of Total Variation (\ref{ringstv}) and Dictionary Learning (\ref{dltomo}), the optimization process can be sped up by a precondition process.
It is well-known that in the back-projected slices, the low spatial frequencies are overrepresented with respect to high frequencies.
In our case, this leads to an ill-conditioned reconstruction problem. 
Therefore, a ramp filter is applied in the Fourier domain to give the proper weight to low and high frequencies.

The filtering is done before back-projection in every iteration, using a discretized version of the high pass ramp filter \cite{murrell}. If there is only one iteration, the optimization then reduces to the standard filtered back-projection. This multiplication by a ramp filter can be seen as a preconditioner transforming an ellipsoidal (ill-posed) problem into a spherical problem, thus fastening the rate of convergence.

The fidelity term $f$ then becomes
\begin{equation}
f(\vec{x}, \vec{r}) = \norm{C \left( \vec{y} - P\vec{x} - \vec{r} \right)}_2^2
\end{equation}
where $C$ is the preconditioner. The gradient is
\begin{equation}
\nabla_{\vec{x},\vec{r}} f = \pmat{\left( C P\right)^T (P\vec{x} - \vec{y} +\vec{r} ) \\ C^T (P\vec{x}-\vec{y}+\vec{r}) }
\end{equation}
Here the operator $\left( C P\right)^T = P^T C^T$ is the filtered back-projection ; so the preconditioner is identified with the high-pass filtering process. We notice that the gradient with respect to the rings variables should also be filtered.


On the other hand, adding rings variables in the functional modifies the condition number of the problem. 
In order to prevent the reconstruction problem to be ill-posed, we introduced a multiplicative coefficient $\alpha$ doing a variable substitution : $S \leftarrow P\vec{x} + \vec{r} - \vec{y}$ becomes $S \leftarrow P\vec{x} + \alpha\vec{r} - \vec{y}$. 
Choosing $\alpha$ enables to pre-condition the problem with respect to the ring variables, at the expense of an additional parameter.

\section{Results}\textbf{\\}
We present here some results for both simulated and real data, and compare our method with two mainstream techniques of rings correction : sinogram pre-processing based on Fourier-Wavelet de-striping \cite{munch09} and image correction using polar coordinates transformation \cite{prell09}.

\subsection{Simulated data}
We use the standard test image ``Lena" containing both smooth components and texture details.
The tests are divided in increasing levels of difficulty for rings removal methods.
First, constant lines are added in the sinogram. These lines independent from the projection angle give raise to rings artifacts in the reconstructed slice.
Since the spurious lines have a constant value, they should be well handled by pre-processing techniques.

Then, lines with variable intensity are added to the sinogram. This makes the correction more difficult for pre-processing techniques, especially if the lines have sharp variations (i.e high frequencies components).

Finally, ring-shaped features are added in the phantom before adding spurious lines in the sinogram.
This case is more challenging because correction methods should not remove any feature coming from the phantom (they belong to the ``true" image), while actually removing rings coming from the sinogram (they come from a flaw in the experimental setup).

The reference sinogram pre-processing technique is the wavelet-Fourier filtering \cite{munch09}. This methods first compute the wavelet decomposition at a level $L$ of the sinogram.
The vertical detail coefficients $V_i$ at level $i \in [[1,L]]$ emphasize the spurious lines that give raise to rings artifacts. In these coefficients, a spurious line is nearly constant along the projection angle, thus, it has only low frequencies in the Fourier domain. Filtering these few low frequencies in the Fourier domain enable to suppress the line after taking the inverse Fourier transform. The filter used is a high-pass Gaussian filter whose standard deviation $\sigma$ tunes the bandwidth. Then, the sinogram is reconstructed from these filtered wavelet coefficients. The Matlab implementation of this method is provided in the author's article.
In the tests, $\sigma$ denotes the standard deviation of the Gaussian filter and $L$ is the number of levels of the wavelet decomposition.

The image correction technique used here is Rings Correction in Polar Coordinates (RCP) \cite{prell09}. It transforms the image into polar coordinates and performs a low-pass filtering in the radial direction. The filtered image is then subtracted from the original image, and a threshold is applied to ignore non-artifact structures. The result is filtered in the azimuthal direction. After a transformation into Cartesian coordinates, the image should only contain rings artifacts ; these are subtracted from the original image. A C++ implementation can be found at \cite{blair14}. 
In the tests, the thresholding parameters are set so that all the image pixels can be considered as possibly part of an artifact.
The important remaining parameters are the maximum ring width $W$, and the maximum angular arc $\theta_0$ we expect the rings to have.

\begin{figure}\label{Fig:directlines}
\includegraphics[width=0.4\columnwidth,height=0.4\columnwidth]{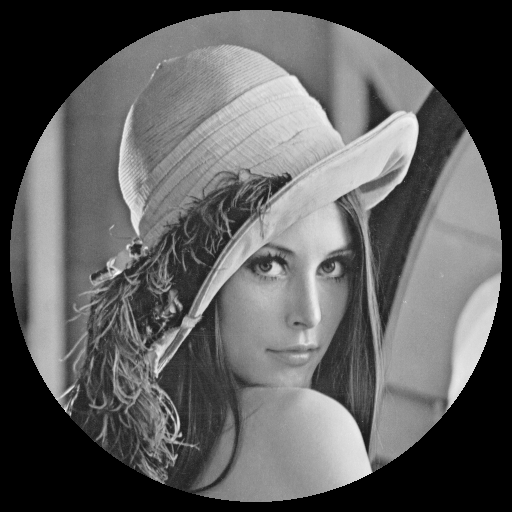}
(a)
\hfill
\includegraphics[width=0.4\columnwidth,height=0.4\columnwidth]{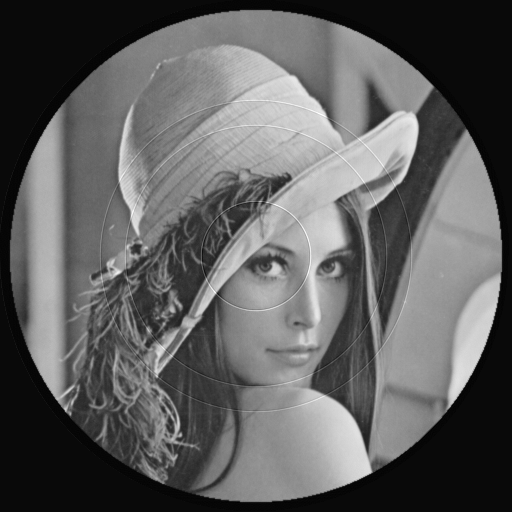}
(b)
\vfill
\includegraphics[width=0.4\columnwidth,height=0.4\columnwidth]{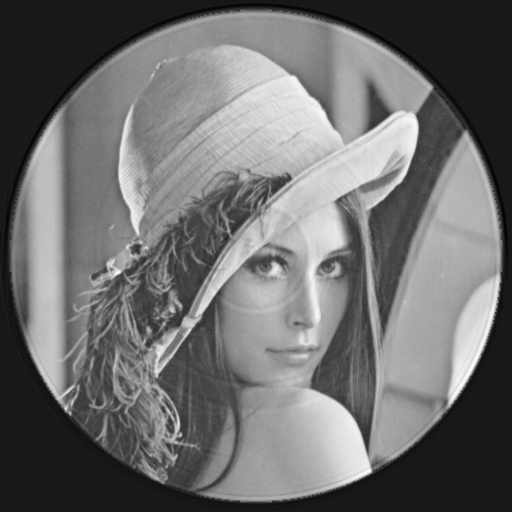}
(c)
\hfill
\includegraphics[width=0.476\columnwidth,height=0.4\columnwidth]{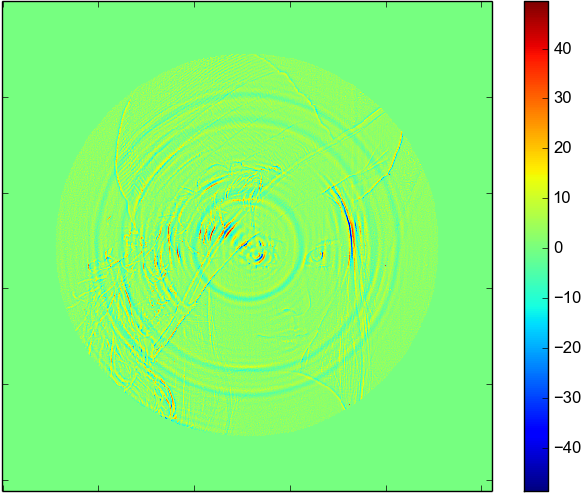}
(d)
\vfill
\includegraphics[width=0.4\columnwidth,height=0.4\columnwidth]{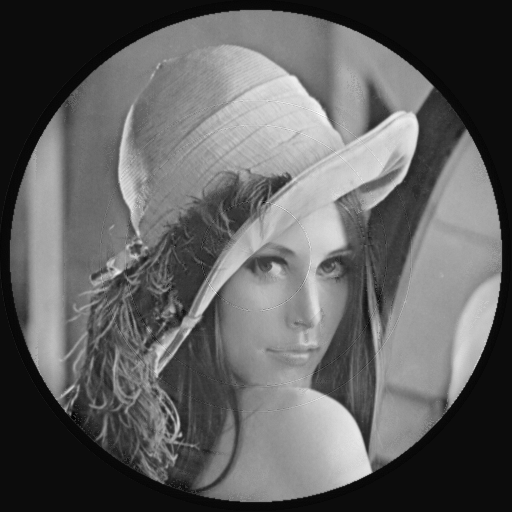}
(e)
\hfill
\includegraphics[width=0.476\columnwidth,height=0.4\columnwidth]{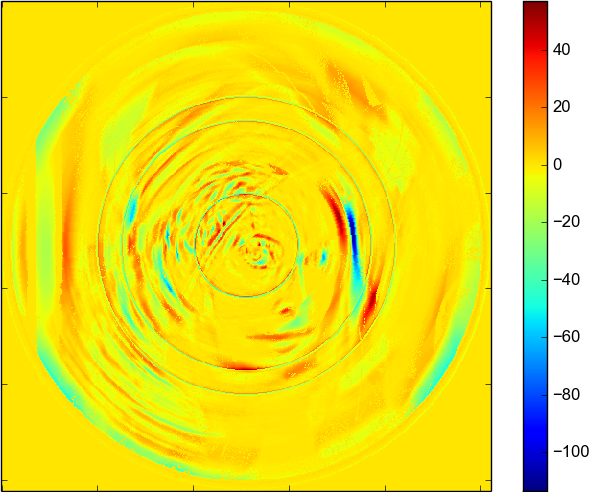}
(f)
\vfill
\includegraphics[width=0.4\columnwidth,height=0.4\columnwidth]{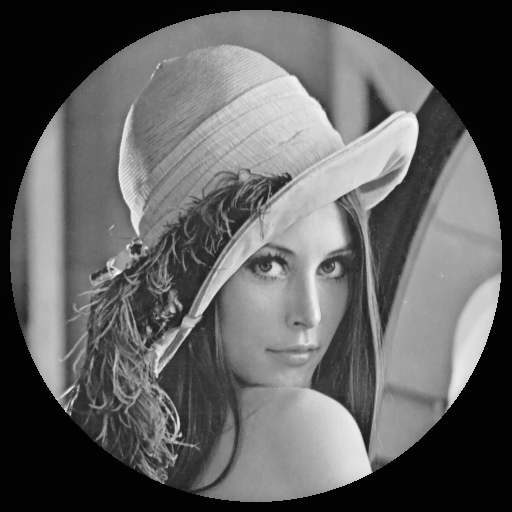}
(g)
\hfill
\includegraphics[width=0.476\columnwidth,height=0.4\columnwidth]{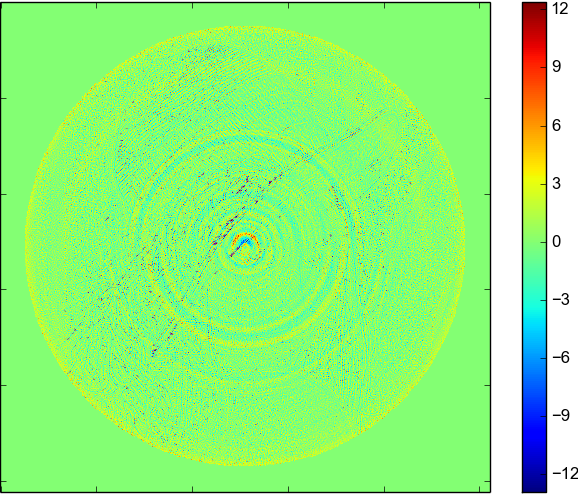}
(h)
\caption{
First test case.\\
(a) Original phantom.
(b) Result of filtered backprojection after adding constant lines in the sinogram.
(c) Image back-projected after applying the Munch \textit{et al.} de-striper algorithm with $\sigma = 3.5$, $L = 2$ and the ``Daubechies 15" wavelet.
(d) Difference between the phantom and the corrected image. The PSNR is 26.6.
(e) Result of the correction with the RCP technique with $W = 10$ and $\theta_0 = 10$.
(f) Difference between the phantom and the corrected image. The PSNR is 29.6.
(g) Result of the reconstruction using the Total Variation regularization in PyHST2, with parameters $\beta = 0.5$, $\beta_r = 0.05$.
(h) Difference between the phantom and the corrected image. The PSNR is 39.0.
}
\end{figure}

Figure \ref{Fig:directlines} shows the results for the firsts test case.
The rings are reduced by the sinogram pre-processing technique (\ref{Fig:directlines}.c), but they do not totally disappear. Besides, additional artifacts appear after the correction. The parameters of this method -- wavelet name, decomposition levels and filter bandwidth -- depend on the image type and the ring width/intensity. In particular, choosing a wavelet with many vanishing moments like in \cite{munch09} does not always yield to better results.
The RCP performs slightly better (\ref{Fig:directlines}.e) ; a strong artifact is added to the right but the result is qualitatively better.
The Total Variation regularization entirely removes the rings (\ref{Fig:directlines}.g).

\begin{figure}\label{Fig:nclines}
\includegraphics[width=0.4\columnwidth,height=0.4\columnwidth]{phantomLena.png}
(a)
\hfill
\includegraphics[width=0.4\columnwidth,height=0.4\columnwidth]{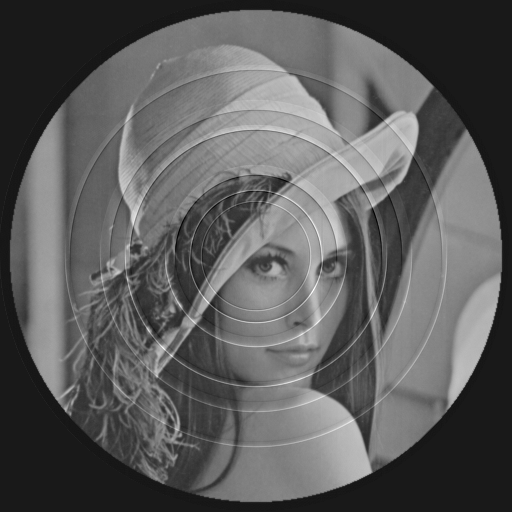}
(b)
\vfill
\includegraphics[width=0.4\columnwidth,height=0.4\columnwidth]{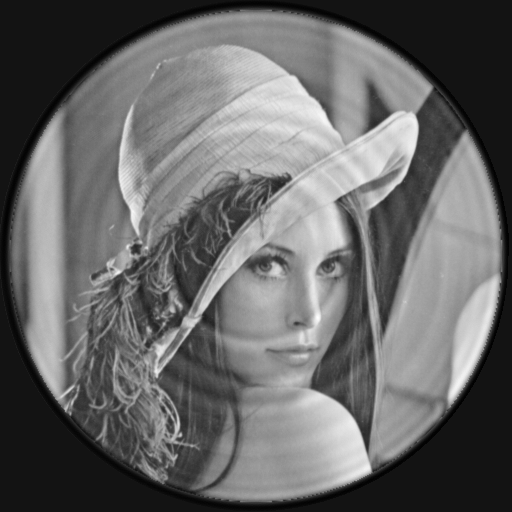}
(c)
\hfill
\includegraphics[width=0.476\columnwidth,height=0.4\columnwidth]{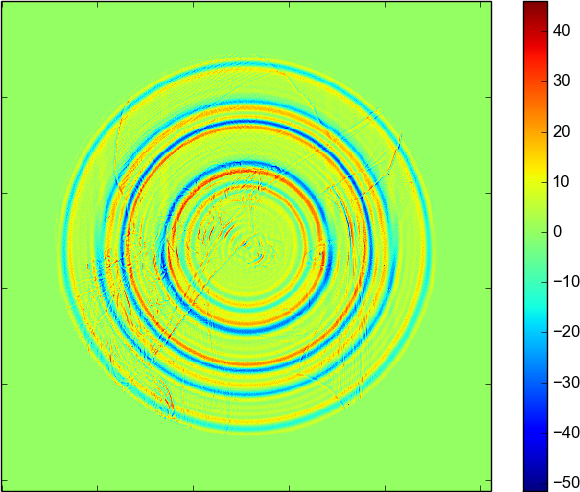}
(d)
\vfill
\includegraphics[width=0.4\columnwidth,height=0.4\columnwidth]{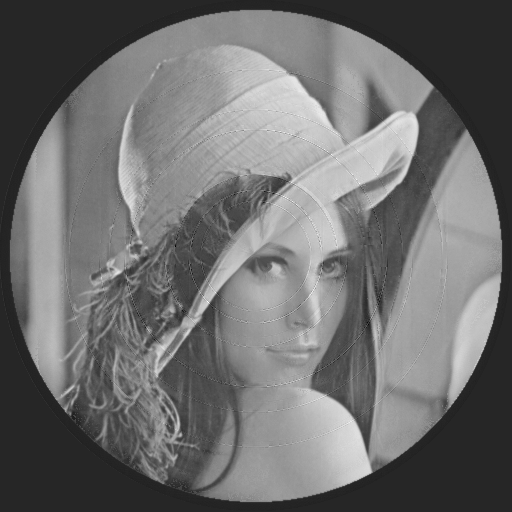}
(e)
\hfill
\includegraphics[width=0.476\columnwidth,height=0.4\columnwidth]{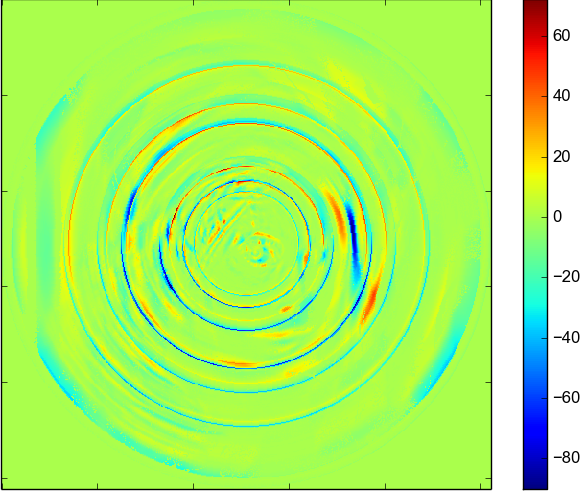}
(f)
\vfill
\includegraphics[width=0.4\columnwidth,height=0.4\columnwidth]{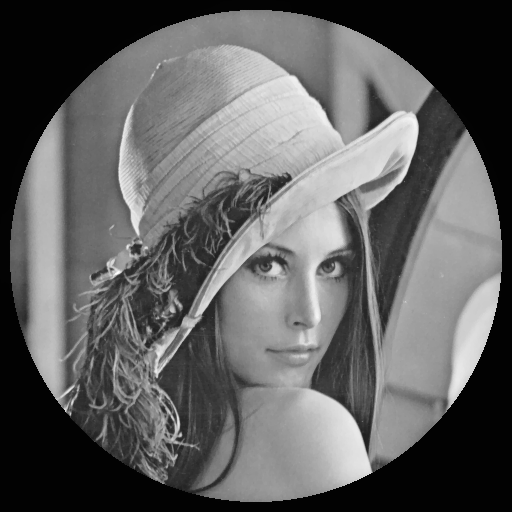}
(g)
\hfill
\includegraphics[width=0.476\columnwidth,height=0.4\columnwidth]{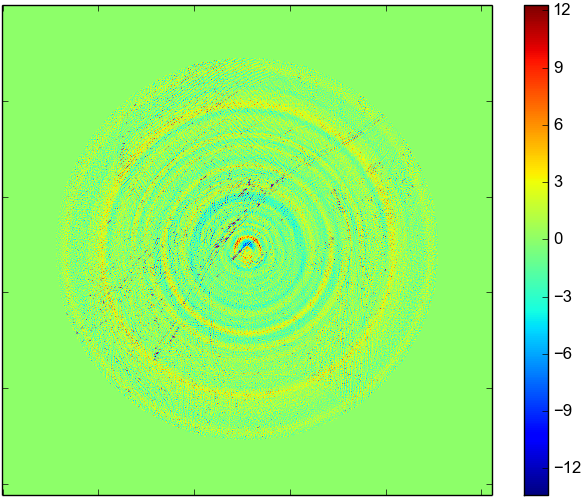}
(h)

\caption{
Second test case.\\
(a) Original phantom.
(b) Result of filtered backprojection after adding lines of variable width and intensity in the sinogram.
(c) Image back-projected after applying the Munch \textit{et al.} de-striper algorithm, with $\sigma = 1.5$, $L = 2$ and the ``Daubechies 15" wavelet.
(d) Difference between the phantom and the corrected image. The PSNR is 29.2.
(e) Result of the correction using the RCP technique with $W = 10$ and $\theta_0 = 10$.
(f) Difference between the phantom and the corrected image. The PSNR is 25.1.
(g) Result of the reconstruction using the Total Variation regularization in PyHST2, with parameters $\beta = 0.5$, $\beta_r = 0.05$. 
(h) Difference between the phantom and the corrected image. The PSNR is 39.4.
}
\end{figure}

Figure \ref{Fig:nclines} shows the results for the second test case.
The sinogram filtering adds many spurious rings (\ref{Fig:nclines}.c). 
The RCP technique removes most of the rings, but a small rings details remain. Again, the difference image shows that strong artifacts are added (\ref{Fig:nclines}.f) even if they are not obviously perceptible on the reconstructed image. 
The Total Variation regularization (\ref{Fig:nclines}.g) entirely removes the rings artifacts ; the result is qualitatively very close to the original phantom.

\begin{figure}\label{Fig:ringsvars}
\includegraphics[width=0.7\columnwidth]{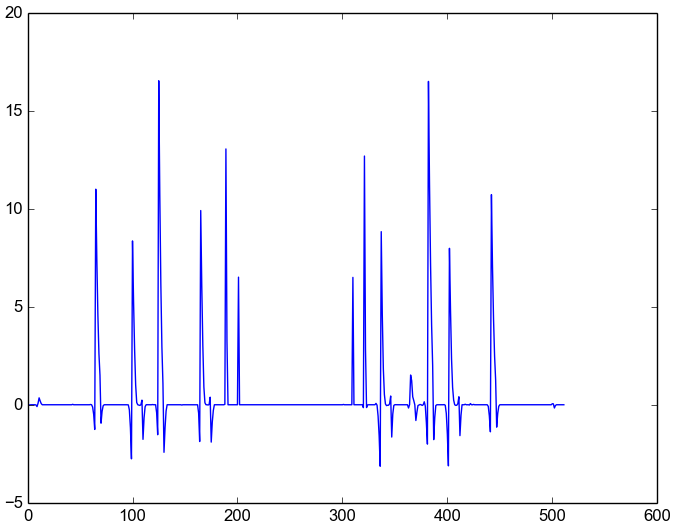}
\caption{
Vector of rings variables for second test case (Figure \ref{Fig:nclines}.b).
The horizontal axis goes from zero to the number of bins of the detector -- that is, in this simulated case, $512$ for the $512\times 512$ test image.
}
\end{figure}

A plot of the rings variables during the total variation reconstruction shows that the rings are actually captured by the ring vector $\vec{r}$.
The six peaks representing the detected lines in the sinogram are clearly visible in Figure \ref{Fig:ringsvars}.

\begin{figure}\label{Fig:disk}
\includegraphics[width=0.4\columnwidth,height=0.4\columnwidth]{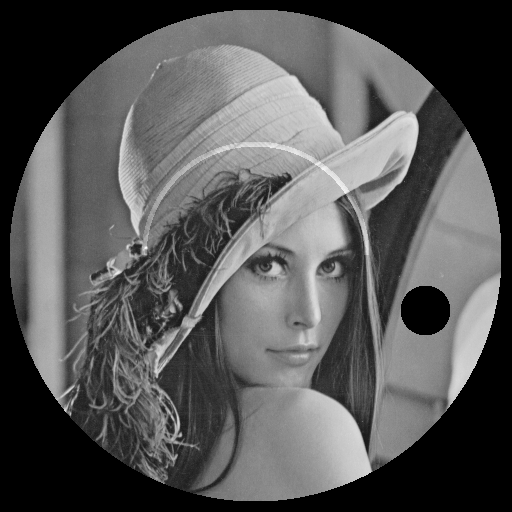}
(a)
\hfill
\includegraphics[width=0.4\columnwidth,height=0.4\columnwidth]{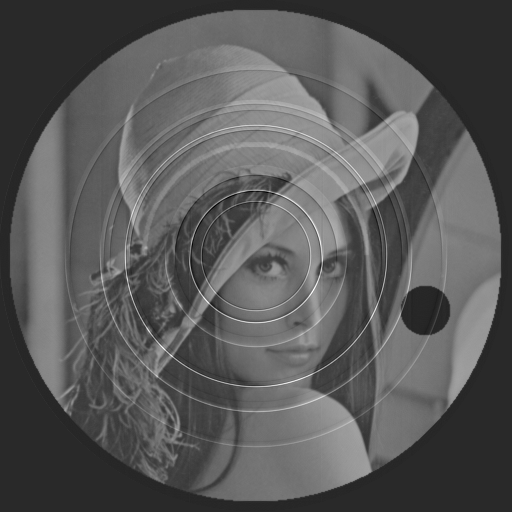}
(b)
\vfill
\includegraphics[width=0.4\columnwidth,height=0.4\columnwidth]{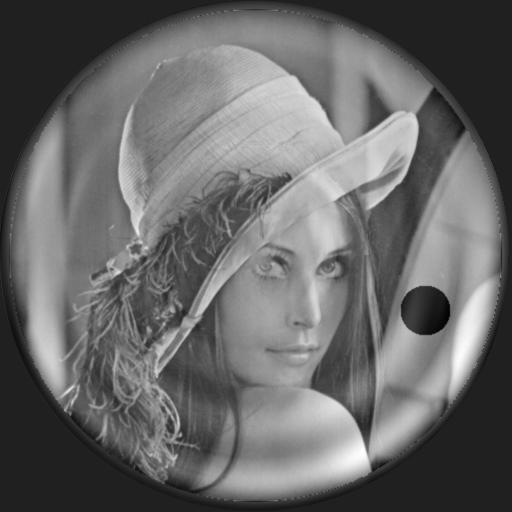}
(c)
\hfill
\includegraphics[width=0.476\columnwidth,height=0.4\columnwidth]{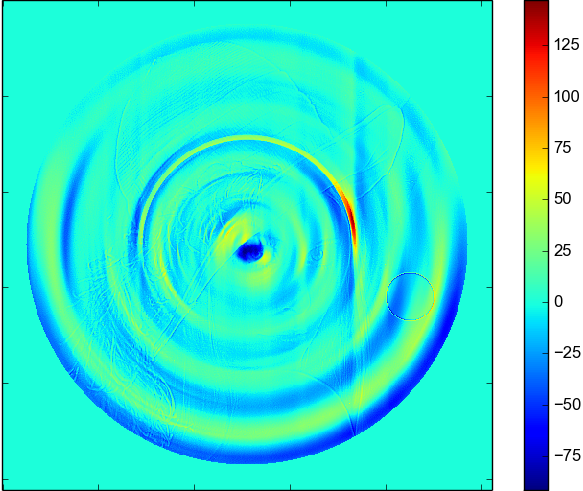}
(d)
\vfill
\includegraphics[width=0.4\columnwidth,height=0.4\columnwidth]{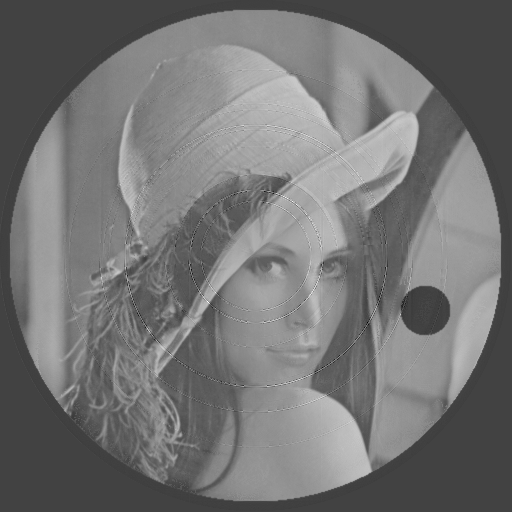}
(e)
\hfill
\includegraphics[width=0.476\columnwidth,height=0.4\columnwidth]{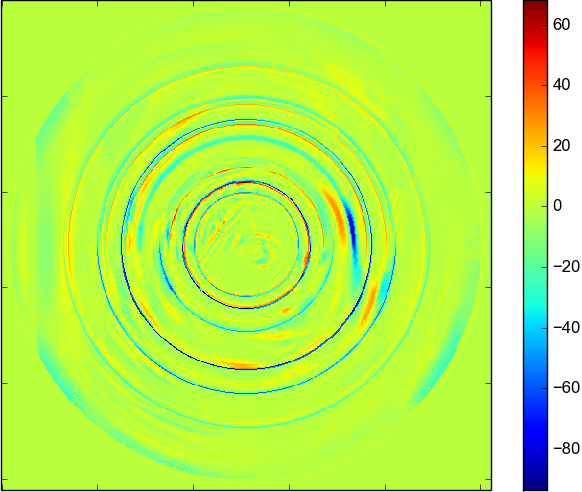}
(f)
\vfill
\includegraphics[width=0.4\columnwidth,height=0.4\columnwidth]{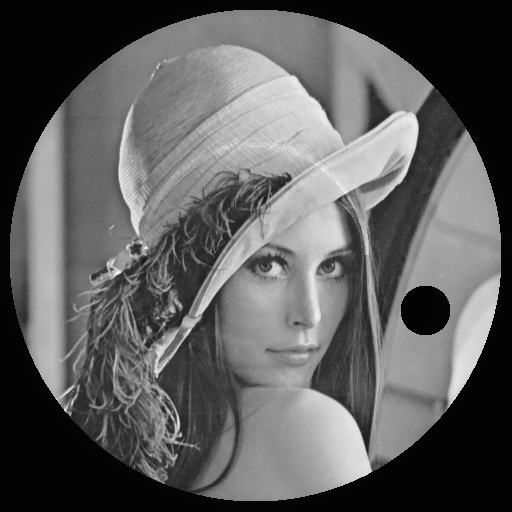}
(g)
\hfill
\includegraphics[width=0.476\columnwidth,height=0.4\columnwidth]{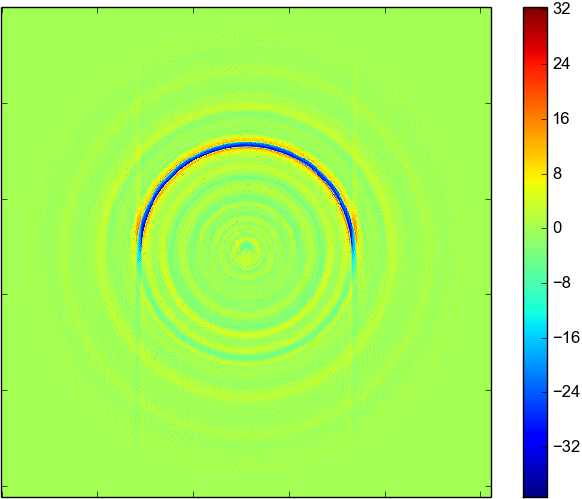}
(h)
\caption{
Third test case.\\
(a) Original phantom.
(b) Result of filtered backprojection after adding lines of variable width and intensity in the sinogram.
(c) Image back-projected after applying the Munch \textit{et al.} de-striper algorithm, with $\sigma = 2.5$, $L = 5$ and the ``Daubechies 20" wavelet.
(d) Difference between the phantom and the corrected image. The PSNR is 23.2.
(e) Result of the correction using the RCP technique, with $W = 10$ and $\theta_0 = 10$.
(f) Difference between the phantom and the corrected image. The PSNR is 21.2
(g) Result of the reconstruction using the Total Variation regularization in PyHST2, with parameters $\beta = 0.5$, $\beta_r = 0.05$. 
(h) Difference between the phantom and the corrected image. The PSNR is 29.9.
}
\end{figure}

Figure \ref{Fig:disk} shows the results for the third test case.
Here two features are added to the original phantom (\ref{Fig:disk}.a) : a black disk and a semi-circular ``ring". These features are part of the phantom, they should not be filtered by rings correction techniques.
Lines added to the sinogram are not constant along the projection angle, and their width can be several pixels. This leads to a back-projected image (\ref{Fig:disk}.b) with large rings.
The RCP technique (\ref{Fig:disk}.e) is more efficient than the sinogram preprocessing (\ref{Fig:disk}.c). The Total Variation rings removal slightly impacts the semi-circular white feature while preserving the black disk.

Beside the visual aspect of the corrected image, we use the Peak Signal to Noise Ratio (PSNR) as a measure of the correction quality :
\begin{equation}
\begin{aligned}
PSNR(I_1, I_2) = 10\log_{10} \left( \dfrac{2^8}{MSE(I_1, I_2)} \right) \\
MSE(I_1, I_2) = \dfrac{1}{N^2}\sum_{i=0}^N \sum_{j=0}^N \left( I_1[i,j] - I_2[i,j] \right)^2
\end{aligned}
\end{equation}
where $MSE(I_1, I_2)$ is the mean square error between the two images $I_1$ and $I_2$ of dimensions $N \times N$ and bit depth of $8$ bits.
Although PSNR gives a score to the overall similarity between the corrected image and the original phantom, it is inconsistent with the eye perception of quality.
For example, RCP performed better than sinogram filtering in these tests, but got a lowest PSNR for cases 2 and 3. The structural similarity index gives the same kind of results.

We were surprised that the sinogram pre-processing technique performed poorly on these simulated cases. 
The reason is probably that the intensity of the lines added to the sinogram is too large for this method.
The technique presented in \cite{munch09} only filters the wavelet detail coefficients : if the sinogram line features are too large, they are captured by the approximation coefficients rather than the detail coefficients.
By trying with lines of smallest amplitude, the Fourier-Wavelet method actually worked without adding big artifacts in the reconstructed image.
For our particular experimental cases, however, the sinogram unwanted lines have non-negligible amplitudes, making this method ineffective.

\subsection{Experimental data}
We give here some results for reconstructions performed on real data. The samples were kindly provided by the ESRF beamline ID19.
\subsubsection{Syntactic foam}
The reconstruction technique was used on a syntactic foam sample.
In this case, the rings are ``large" in the extent that the radius difference between the exterior and the interior of the ring is several pixels.
This means that the spurious lines in the sinogram have several pixel of width along the detector bins axis, forming ``bands".
However, the intensities of the lines forming a band has too many variations to be efficiently filtered by sinogram pre-processing techniques.
This case is also difficult for slice-correction algorithms which detect circular features, since the sample itself have circular features we do not want to remove.

\begin{figure}\label{Fig:mousse}
\includegraphics[width=0.4\columnwidth,height=0.4\columnwidth]{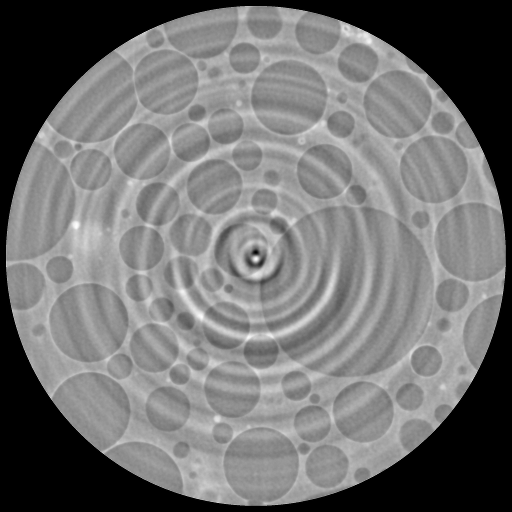}
(a)
\hfill
\includegraphics[width=0.4\columnwidth,height=0.4\columnwidth]{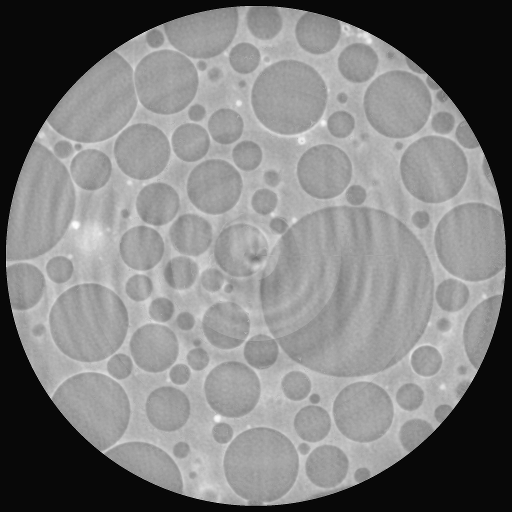}
(b)
\vfill
\includegraphics[width=0.4\columnwidth,height=0.4\columnwidth]{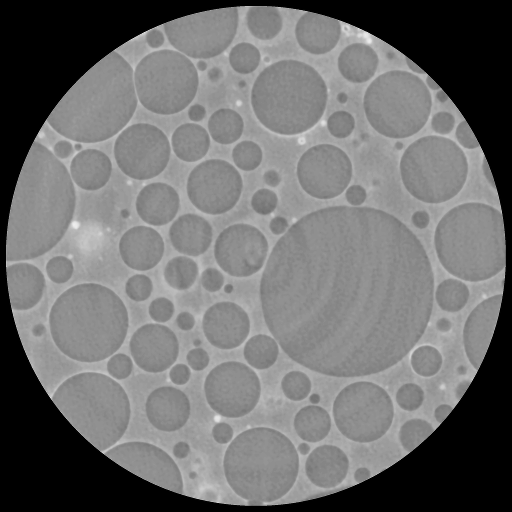}
(c)
\caption{
Syntactic foam sample tomography acquired at ESRF ID19, with energy of 19 keV and pixel size of $0.28 \mu m$.
(a) Filtered back-projection.
(b) Correction with RCP technique, using the parameters $W = 60$ and $\theta_0 = 60$.
(c) Reconstruction with Dictionary Learning technique in PyHST2, using the parameters $\beta_r = 0.1$, $\beta_{\text{DL}} = 0.035$ and $\rho = 20$.
}
\end{figure}

\subsubsection{Rhynie chert}
We apply the reconstruction on a rhynie chert sample.
This situation is almost the opposite of the previous case : the rings artifacts have a small intensity in the reconstructed slice, and the sample borders form a nearly circular polygonal shape.
These border have a huge amplitude with respect to the rest of the sample, and the transition between the border and the interior/exterior is very sharp.
Thus, slice correction techniques would try to remove the borders before any other feature in the slice depending on the thresholding parameters.

\begin{figure}\label{Fig:fossil}
\includegraphics[width=0.4\columnwidth,height=0.4\columnwidth]{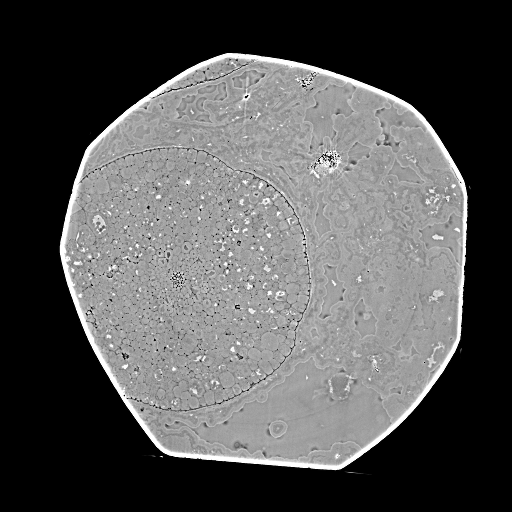}
(a)
\hfill
\includegraphics[width=0.4\columnwidth,height=0.4\columnwidth]{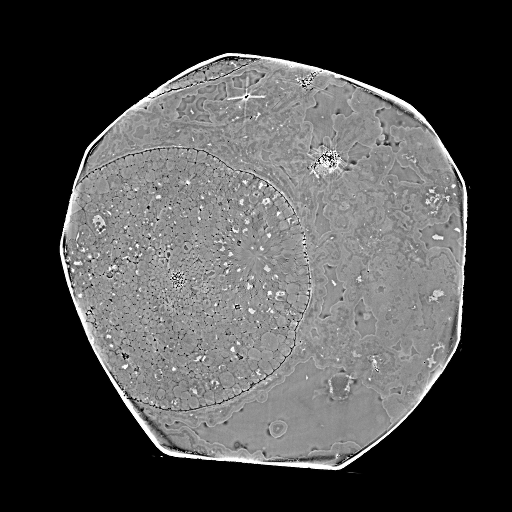}
(b)
\vfill
\includegraphics[width=0.4\columnwidth,height=0.4\columnwidth]{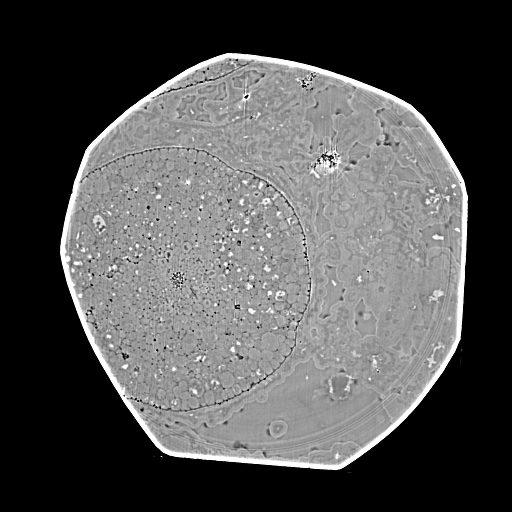}
(c)
\hfill
\includegraphics[width=0.4\columnwidth,height=0.4\columnwidth]{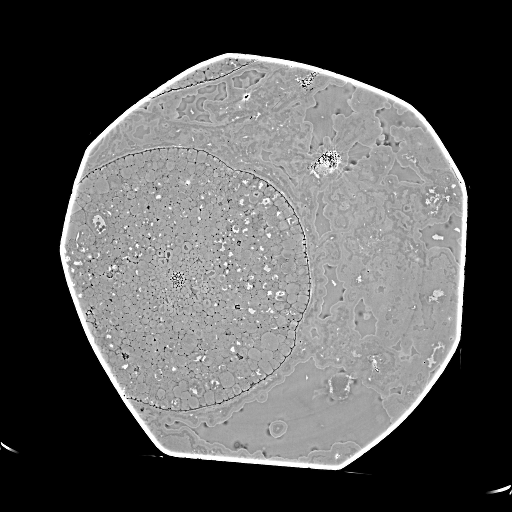}
(d)
\caption{
Rhynie chert sample tomography acquired at ESRF ID19, with energy of $17.6$ keV and pixel size of $1.52 \mu m$.
(a) Filtered back-projection with the correct rotation axis.
(b) Correction with the RCP technique, using the parameters $W = 10$ and $\theta_0 = 10$.
(c) Reconstruction with the Total Variation regularization using the incorrect rotation axis.
(d) Reconstruction with the Total Variation regularization using the correct rotation axis. The parameters were $\beta = 3\E{-3}$ and $\beta_{r} = 3\E{-4}$.
}
\end{figure}

We realized that the rings correction was difficult for the total variation reconstruction : the procedure added rings tangent to one of the slice borders.
It turned out that the problem was due to the rotation center for (back)projection improperly set, leading to accumulating errors in the iterative reconstruction.
Indeed, total variation and dictionary learning reconstruction require to compute the projection for the functional, and the back-projection for the functional gradient. 
If the rotation center for these operations is not the same that the one used for actually rotate the sample, slight errors appear in the (back)projection ; these error accumulate with the number of iterations and take the form of circular features (Figure \ref{Fig:fossil}.c).

After setting the correct rotation center, we were able to remove the ring artifacts (Figure \ref{Fig:fossil}.d), especially the one near the center of Figure \ref{Fig:fossil}.a. In this case, the RCP technique did quite well (Figure \ref{Fig:fossil}.b).

\subsection{Execution time and convergence rate}
In this section we measure the execution time required to obtain an acceptable reconstruction. All the tests are performed on a machine with an \verb|Intel Xeon CPU E5-1607 v2 @ 3.00GHz| processor and a \verb|GeForce GTX 750 Ti| graphic card.

\begin{figure}\label{Fig:bench1}
\includegraphics[width=0.8\columnwidth]{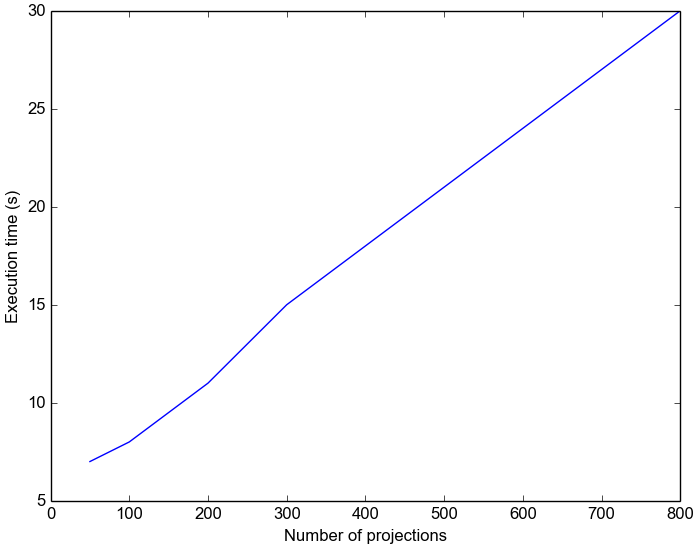}
\caption{
Execution time as a function of the number of projection. The image used is the $512 \times 512$ test image ``Lena" corrupted with the rings presented in the second test case.
}
\end{figure}
We measured that the execution time is the same with rings correction and without rings correction, which is coherent since the ring correction is part of the functional.
Hence, the execution time gives an insight of the time needed by Compressed Sensing reconstruction, but it is not the best choice to measure the influence of rings correction on the overall reconstruction time.
We now focus on the number of iterations required to converge to an acceptable solution. The cost of a single iteration depends on the number of projections, as it can be guessed with Figure \ref{Fig:bench1}.

To measure the convergence rate, we use the values of the objective function. For Total Variation reconstruction, the objective function is given by \eqref{ringstv}, it includes both the fidelity term (Euclidean distance) and the regularization term (L1 norm of the image gradient). One can notice that the energy is not a monotonic function of the number of iterations, which can be seen as a drawback of FISTA with respect to \nopagebreak ISTA.

\begin{figure}\label{Fig:bench2}
\includegraphics[width=0.8\columnwidth]{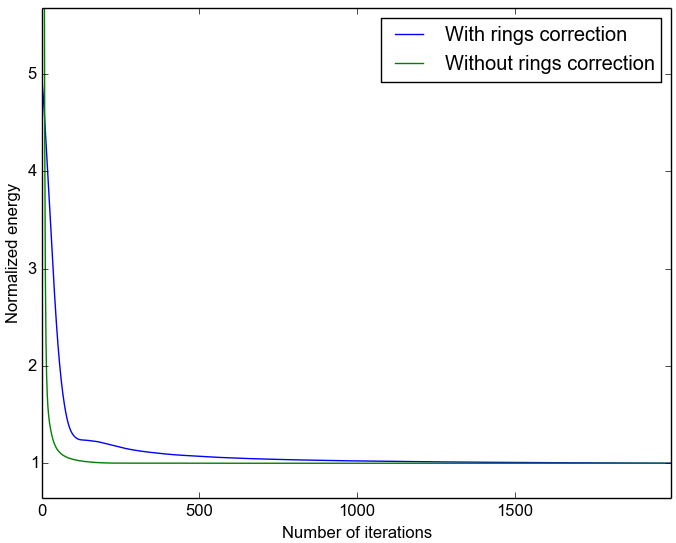}
(a)
\hfill
\includegraphics[width=0.8\columnwidth]{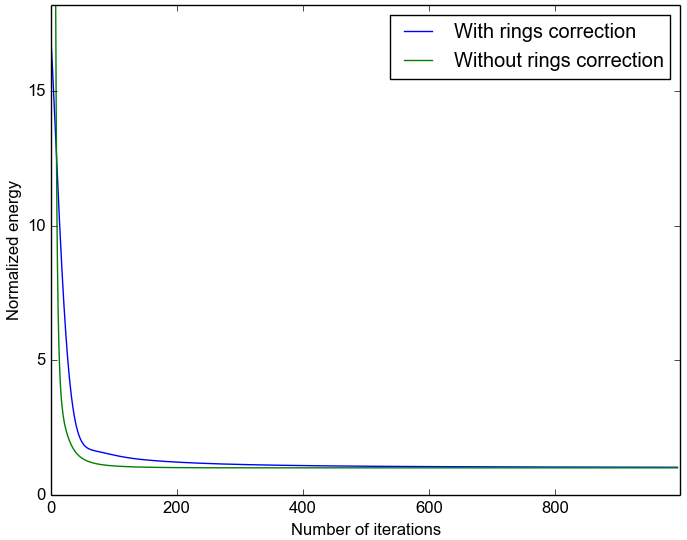}
(b)
\caption{
Energy as a function of the number of iterations for the Total Variation tomographic reconstruction. 
The energy is normalized by the energy of the last iteration in order to have the same scale in the two cases.
The image used is the $512 \times 512$ test image ``Lena" corrupted with the rings presented in the second test case.
\textbf{(a)} Evolution of energy with 800 projections.
\textbf{(b)} Evolution of energy with 200 projections.
}
\end{figure}

With rings correction, the reconstruction process takes more iterations to converge, as illustrated in Figure \ref{Fig:bench2}.
For simulated and real data, it turned out that a satisfactory reconstruction can be achieved with less than $1000$ iterations without rings correction. 
When the rings correction is activated, it takes about $2000$ iteration to correctly remove the ring artifacts.
Thus, while the rings correction has no additional cost per iteration, it takes nevertheless more iterations to converge to an image with removed ring artifacts.
The ``energy transfer" between the fidelity term $\norm{\vec{y} - (P \vec{x} + \vec{r})}_2^2$ and the L1 norm of the rings $\norm{\vec{r}}_1$ is actually quite slow.

The convergence rate also depends on the number of projections. Figure \ref{Fig:bench2} shows that the reconstruction process converges in 500 iterations (400 with rings correction) for 200 projections, when it takes about 2000 iterations (1500 without rings correction) for 800 projections.
It is due to the fact that 
the weight of fidelity term virtually increases as the number of projection increases. To counter-weight this, one has to increase the penalty term $\beta$ of the regularization, which makes the energy transfer between the fidelity term and the ring variables a little faster.

The reconstruction quality is still excellent with four times less projections -- which is a crucial interest of Compressed Sensing tomographic reconstruction.
Therefore, this method makes sense rather for low-dose tomography when there are few projections ; and for now it is quite expensive to use it only for rings correction when a reconstructed volume is already available.


\subsection{Conclusions}
We presented a new way to correct the rings artifacts that appear in tomographic reconstruction.
Compressed sensing tomographic reconstruction is a promising technique which enables to obtain a slice of good quality with fewer projections.
Including the rings artifacts correction in the iterative reconstruction process has shown to be efficient while requiring no extra pre or post-processing steps.
Besides, additional artifacts are less likely to appear thanks to the regularization. 
This method can be adapted to any compressed sensing approach, since the only things to do are modifying the functional and the iterative correction step accordingly.

In a further work, we would like to improve the convergence rate of the rings correction in order to make it more attractive. We also would like to extend this method to lines that are not constant along the projection angle in the sinogram, in order to cover more general and difficult cases.




\ack{\textbf{Acknowledgements}} 
We thank Elodie Boller and Paul Tafforeau from the ESRF beamline ID19 for having kindly provided experimental samples.
We also thank Michel Desvignes from Gipsa-Lab, Grenoble, for his reviews and advices.


\bibliographystyle{iucr} 
\bibliography{biblio}






\end{document}